\documentclass[showpacs,amsmath,amssymb,twocolumn,aps]{revtex4}

\usepackage{graphicx}
\usepackage{epsfig}
\usepackage{pstricks}
\usepackage{psfrag}
\usepackage{dcolumn}
\usepackage{bm}

 \def\comment#1{}
 \def\mn#1{}

\begin{document}

\title{Lindemann Parameters for solid Membranes focused on Carbon Nanotubes}

\author{J\"urgen Dietel}
\affiliation{Institut f\"ur Theoretische Physik,
Freie Universit\"at Berlin, Arnimallee 14, D-14195 Berlin, Germany}
\author{Hagen Kleinert}
\affiliation{Institut f\"ur Theoretische Physik,
Freie Universit\"at Berlin, Arnimallee 14, D-14195 Berlin, Germany}
\affiliation{ICRANeT, Piazzale della Repubblica 1, 10 -65122, Pescara, Italy}
\date{Received \today}

\begin{abstract}
Temperature fluctuations
 in the normal direction
of planar crystals
 such as graphene are quite violent and
may be expected to influence
strongly their melting
properties. In
particular, they will modify
the Lindemann melting criterium.
We calculate this modification
in a self-consistent
Born approximation.
The result
is applied to graphene and
its wrapped version represented by
single-walled carbon nanotubes (SWNTs). It is found that
the out-of-plane fluctuations dominate over the
in-plane fluctuations. This makes strong restrictions to possible
Lindemann parameters.
Astonishing we find that these large out-of-plane fluctuations
have only a small  influence
upon the melting temperature.
\end{abstract}


\maketitle

\section{Introduction}
The production of macroscopic two-dimensional (2D) graphene sheets by
mechanical cleavaging \cite{Nosolev1} has demonstrated that
free-standing or suspended 2D crystals can exist despite
large positional fluctuations in two dimensions.
Since then,
a variety of other free-standing 2D crystallites have been
 prepared \cite{Nosolev2}.
Wrapped versions of the 2D free-standing graphene
had been found much earlier in 1991 \cite{Iijima1}.
Recent observations  \cite{Meyer1} have
confirmend the theoretical expectation
that freely suspended
graphene sheets are strongly
undulated and behave more like solid membranes
than two dimensional (2D) crystals \cite{Nelson1}. The undulations are
a consequence of the thermal fluctuations of the membrane.
In this paper,
we calculate these fluctuations quantitatively
and discuss their implications upon the melting
properties such as Lindemann parameter and melting temperature.
 \cite{Meyer1}
The results will be compared with
corresponding 2D crystals.

The easiest way to estimate the melting temperature of a
three  dimensional (3D)
crystal
is based on the Lindemann criterium \cite{Lindemann1}.
According to it, a 3D crystal starts to melt
when the
square root of the
thermal expectation value of
lattice site elongations
$ \sigma _i\equiv  \sqrt{ \langle {\bf u}^2_{i} \rangle_T \/}$
exceeds a certain fraction of the
nearest-neighbor lattice distance $ a $, usually
around $ 0.1-0.15 $ \cite{GFCM2}.
Above the melting temperature, the shear modulus of the
lattice vanishes leading
to a
divergence
in the displacement
fluctuations typical for the liquid state.

In two dimensions (2D) this criterium is no longer
applicable since the
displacement fluctuations are always logarithmically divergent,
reflecting the fact
that after a long time,
a 2D crystal migrates through the entire 2D-space.
There exists, however,
a simple modification \cite{Bedanov1}.
Instead of $  \langle {\bf u}^2_{i} \rangle_T$
 one may use
the
finite
cumulants
$\langle {\bf r}^2_{ij} \rangle_T - \langle {\bf r}_{ij} \rangle_T^2 $,
where $ {\bf r}_{ij} $ is the difference vector between the
atoms associated with the nearest-neighbor
lattice sites $ i $ and $ j $. This leads to
a modified
Lindemann number
\begin{equation}
{\cal L}^{s,2D}_1 =  \frac{1}{|{\cal N}_1|}
\sum\limits_{i, j \in {\cal N}_1} \frac{\sqrt{
\langle {\bf r}^2_{ij} \rangle_T - \langle {\bf r}_{ij} \rangle_T^2  }}{
a}      \,.      \label{10}
\end{equation}
Here
$ {\cal N}_1 $ denotes the
set of all nearest-neighbor lattice pairs and  $ |{\cal N}_1| $
their number.
For the Lennard-Jones
and Wigner lattices,
Bedanov {\it et al.}   \cite{Bedanov1} found
 by computer simulations
values of
$ {\cal L}^{s,2D}_1 \approx 0.15-0.2 $.
We have derived
the same values
analytically
for a
triangular generalization
\cite{Dietel1} of
a square lattice defect melting model \cite{GFCM2}.

   At this point it is
useful to realize that
a migration problem
and an associated divergence of
$\sigma_i $
exists also in three dimensions
if the system is finite, i.e., for
 3D clusters and polymers. There one defines
 a modified Lindemann number
\begin{equation}
 {\cal L}^c_{1,2} =  \frac{1}{|{\cal N}_{1,2}|}
\sum\limits_{i, j \in {\cal N}_{1,2}}
\frac{\sqrt{
\langle {\bf r}^2_{ij} \rangle_T - \langle {r}_{ij} \rangle_T^2 }}
 { \langle {r}_{ij}  \rangle_T }    \,.        \label{20}
\end{equation}
Here  $ {\cal N}_2 $ is the set of {\em all\/} lattice site pairs
whose number is $ N (N-1)/2 $ where $ N $ is
 the number of atoms in the lattice.
The number $ {\cal L}^c_{1} $ was introduced by
Kaelberer and Etters \cite{Kaelber1},
the number $ {\cal L}^c_{2} $
by  Berry {\it et al.} \cite{Berry1}.
For small clusters,
 $ {\cal L}^c_{1} $ and
$ {\cal L}^c_{2} $ have similar values \cite{Zhou1}
of around
$ {\cal L}^c_{1,2} \approx 0.03-0.05 $ at the melting point.
Above the melting point all modified
Lindemann numbers
increase
considerably, but do not go to infinity
(this being in contrast to $ \sigma _i$).

The main difference between
$ {\cal L}^{s,3D}_1 $ and $  {\cal L}^c_1 $ comes from the
last term in the square root in (\ref{10}) and (\ref{20}).
Whereas
$  \langle {\bf r}_{ij} \rangle_T $ is
the temperature
average of the difference vector of sites $ i $ and $ j $, i.e.
$ {\bf r}_{ij}= (x_{ij}, y_{ij} ,z_{ij}) $, the expectation
value
$  \langle {r}_{ij} \rangle_T $ is the
average value
of the bonding length of sites $ i $ and $ j $, i.e.
$ {r}_{ij}=({x^2_{ij}+ y^2_{ij} + z^2_{ij}})^{1/2} $.
Since ${\bf r}^2_{ij} = {r}_{ij}^2 $ one would expect that
the  3D version of (\ref{10})$, {\cal L}^{s,3D}_1 $,
and $ {\cal L}^{c}_1 $ (\ref{20}) could be equally
useful in determining the melting point.
This is indeed the case for 3D crystals.

In this paper we shall consider
all three
Lindemann numbers
 ${\cal L}^{s,3D}_1 $,
$ {\cal L}^{c}_1 $, and $ {\cal L}^{c}_2 $
as candidates for a melting criterion
for solid membranes
such as graphene lattices or
SWNTs.
It will turn out
that
for these
${\cal L}^{s,3D}_1 $ is unsuitable
 for calculating the melting temperature.
The reason lies in the large out-of-plane fluctuations
of the membrane vary little when crossing the
melting point.
These fluctuations cancel each other in (\ref{20})
since  $ \langle z^2_{ij} \rangle_T \not= 0 $ and
$ {r}_{ij}\approx  ({x^2_{ij}+ y^2_{ij}})^{1/2} +
(1/2)z^2_{ij}/ ({x^2_{ij}+ y^2_{ij}})^{1/2}  $
but
not in  since ${\cal L}^{s,3D}_1 $ since  $ \langle z_{ij} \rangle_T=0 $.

Freely suspended
graphene sheets are always undulated and behave like a solid membrane
\cite{Meyer1}.
Nelson {\it et al.} \cite{Nelson1, Nelson2} have shown that
in-plane fluctuations tend to stabilize a solid membrane
such that a flat phase can exist in spite of
its large 2D fluctuations.
The melting temperature of  (5,5) SWNTs
was determined
by Zhang {\it et al.} \cite{Zhang1} within
numerical simulation to be around $ T_m \approx 5000\,$K,
 in agreement with experimental determinations
\cite{Huang1}.
The value
of the Lindemann number
$ {\cal L}^c_{2} $ was around $ {\cal L}^c_{2} \approx 0.03$ at the
onset of melting defined by
the abrupt
increase of $ {\cal L}^c_{2} $. However, when using
the region of
strong increase of the internal energy  they obtain a range
$ {\cal L}^c_{2} \approx 0.03 \sim 0.07$ from the onset
of melting to
its completion.

The shapes of SWNTs near the melting temperature
are in general
strongly deformed from a pure tube form. This leads to the conclusion
that the 2D nearest-neighbor Lindemann number $ {\cal L}^{s,2D}_1 $
(\ref{10}) is not a useful quantity for a melting criterion.
One rather should use the  Lindemann-like
numbers (\ref{20})
or the 3D form of (\ref{10}) which both respect the 3D rotational
 symmetry of the system. In the following, we shall
first calculate $ {\cal L}^c_{1} $ which
for small clusters
and small supercells
agrees in molecular dynamic simulations
with $ {\cal L}^c_{2} $.
We shall restrict ourselves to the (5,5) SWNT
so that we
can compare our theoretical results with existing simulation data.
We shall find that despite the large vertical fluctuations
of the membranes,
the Lindemann number (\ref{20})
depends
mostly on the
in-plane
fluctuations
and provides us with a valuable melting criterium.
This is not the case for the 3D-version
of the Lindemann number $ {\cal L}^s_{1} $ having its reason in the
fact that the out-of plane fluctuations are even larger than the in-plane
fluctuations for SWNTs and graphene.
Surprisingly,
the melting
temperature of SWNTs is modified only little
by these large out-of-plane fluctuations
at high temperatures.
\mn{is this not related to their deformation?}

\section{Membranes}

The elastic energy of a  solid elastic membrane in the flat phase
is given by \cite{Landau1}
\begin{equation}
 H_{\rm el}=  \int d^2x\left[  \mu u_{ij}^2 + \frac{1}{2} \lambda u_{ii}^2
+ \sigma_{ij} u_{ij}
+ \frac{1}{2} \kappa_0 (\nabla_i \nabla_i f)^2 \right] ,
\label{30}
\end{equation}
 where
\begin{equation}
 u_{ij}= \frac{1}{2} \left(\nabla_i u_j +  \nabla_j u_i +
 \nabla_i f \nabla_j f\right)  \,.
\label{40}
\end{equation}
and $ u_j $ are the lattice displacements in $xy$-plane,
while $ f $ is the
out-of-plane displacement. The constant $ \mu $ is
the shear modulus, and $ \lambda $
is the Lam\'{e}  constant. The last term in
(\ref{40})
with the constant $ \kappa_0 $
 accounts for the
 bending stiffness of the membrane.
The quantity
$ \sigma_{ij} $
is an external stress source
which will help us to calculate (\ref{10}), (\ref{20})
from derivatives of the
partition function with respect to $ \sigma_{ij}$. The line element
on the membrane for small distortions is given by \cite{Landau1}
$ d l'^2= dl^2 +2 u_{ij} dx_i dx_j $, where $ dl $ is
the length of the undistorted
planar surface, which we identify with the equilibrium
lengths $l_{ij}$ between sites $ i $ and $ j $.
Thus we calculate $ {\cal L}^c_{1,2} $ by first inserting  the line element
$ d l' $ in (\ref{20}), and afterwards expanding the resulting expressions
for small displacements.
Thus we obtain
\begin{equation}
 {\cal L}^c_{1,2} =   \sum\limits_{i, j \in {\cal N}_{1,2}} \! \! \! \!
\frac{\sqrt{ \left\langle ( u_{lm} \, a^2 e^{ij}_l e^{ij}_m)^2\right\rangle_T -
\left\langle u_{lm} \, a^2 e^{ij}_l e^{ij}_m\right\rangle_T^2} }
 {|{\cal N}_{1,2}|\quad   a^2}   \,.          \label{50}
\end{equation}
Here $ {\bf e}^{ij} $ are
the unit vectors pointing from
site $ i $ to $ j $.
In deriving (\ref{50})
we used a Taylor
expansion of the elongation differences between two lattice
sites (gradient expansion). This is
 justified for the
small elongation differences
of neighboring atoms occurring
in $ {\cal L}^s_1 $ and
$ {\cal L}^c_1 $
up to the melting regime.
In the atom pairs summed over
in
$ {\cal L}^c_2 $, the approximation is good only for
small clusters or
for small supercells in molecular dynamic
simulations.
For infinite solid membranes,
 this is no longer the case.
The contributions of the
far separate pairs $(i,j)$
cause
a drastic decrease of the Lindemann parameter $ {\cal L}^c_2 $
with the size of the system.
In a 2D crystal, for example,
the widely separated pairs
contribute terms which grow logarithmically
with the separation:
$ \lim_{l_{ij \to \infty}}
\langle {\bf r}_{ij}^2 \rangle_T - \langle {\bf r}_{ij}
\rangle^2_T \sim \ln l_{ij} $.
This
eliminates $ {\cal L}^c_2 $
for
determining the melting point.

In order to calculate (\ref{50}) we first integrate out
the $xy$-lattice displacement
fields $ u_i $ in the partition function, leading to
an effective Hamiltonian $ H= H_{h}+H_{\sigma \sigma}$
with
\begin{align}
&\!\!\! \!H_{h}= \!\int\! d^2 x\!\left\{  \tilde{\mu}
\left[(h_{f}+h_{\sigma})^2 - h^2_{\sigma}\right] +
\frac{1}{2} \kappa_0 (\nabla_i \nabla_i f) \right\}  , \label{60}  \\
&\!\!\!\! H_{\sigma \sigma}\!= \!\int\! d^2 x \! \left[\frac{1-\nu}{4\mu}
\left(\frac{\nabla_i \nabla_j}{\nabla_l \nabla_l}  \sigma_{ ij}\!\!\right)^2 +
\frac{1}{2\mu}
\left(\frac{\nabla_i \nabla^T_j}{\nabla_l \nabla_l} \sigma_{ ij}\!\!\right)^2
 \right] ,\nonumber
\end{align}
and
the energy densities
\begin{align}
& h_{f} = \frac{1}{2}  P^T_{l m}
 ( \nabla_l f \, \nabla_m f )  \,,                        \label{70}  \\
& h_{\sigma} = \frac{1}{2 \mu} P^T_{lm}
\left[ \!1 - \!\delta_{l m} \frac{\nu}{1+ \nu}\left(\! 1 +
\frac{\nabla_l \nabla_l}{(1-\delta_{l k})  \nabla_{k} \nabla_{k} }\! \right)
\! \right]  \sigma_{l m}  \,.  \nonumber
\end{align}
In (\ref{60})
we have used the abbreviation $ \tilde{\mu} \equiv  \mu (1+ \nu)=E/2 $ where
 $E$ is the Young modulus and $ \nu\equiv \lambda/(2 \mu+ \lambda) $
  the Poisson ratio.
The calculation of the energy densities (\ref{70})
is simplified by
the fact
that only the transverse part
$ P^T_{l m} \nabla_l f
\nabla_m f $ with
$  P^T_{l m}= (\delta_{l m} - q_{l} q_{m}/q^2) $
of the out-of-plane fluctuations
is relevant after integrating out the in-plane fields $ u_i $
\cite{Nelson1}. The
transverse projections
lead to a useful
 restriction of the relevant phase space when
calculating Feynman diagrams.

\subsection{Self-consistent Born approximation}

We now treat the Hamiltonian (\ref{60}) within
the self-consistent Born-approximation (SCBA)
corresponding to the Hartree-Fock approximation
for the
eigenfunctions.
Other approximations to the Hamiltonian (\ref{60})
have been used
\cite{Bowick1} to calculate
the universal roughening exponents of the
membrane, for example in Ref.~\onlinecite{Doussal1}
an extension of SCBA.

Denoting the inverse Green function of the $ f $-fluctuations
by $ G^{-1}({\bf k}) =\kappa_0 k^4 + \Sigma(k) $,
we obtain from (\ref{60}) within
the SCBA 
\begin{equation}
\Sigma(k)= \frac{2 \tilde{\mu} k_B T}{(2 \pi)^2}  \int\limits_{\rm BZ}
d^2 q \,
\frac{({\bf k} \times {\bf q})^4}{q^4} \, G({\bf k} + {\bf q})
         \label{80}
\end{equation}
where we take into account only the Fock-part of the SCBA.
It was shown in Ref.~\onlinecite{Nelson2} that the Hartree-terms
do not contribute
for free boundary conditions of the $xy$-elongations of the membrane.
To do the integral in (\ref{80})
we use
a circular Brillouin zone $ k \le k_{\rm BZ} $,
with
$ k_{\rm BZ}=8 \pi/\sqrt{3} a^2 $ for the triangular Bravais
lattice of SWNTs and graphene.  The integral (\ref{80}) can be
carried out exactly for  small $ k $ \cite{Smirnov1}
to
obtain the first two terms in
the expansion
\begin{equation}
 \Sigma(k) = C_T k^3 + \kappa_{\scriptstyle\Sigma}k^4+\dots\, ,
\label{@SIGEX}\end{equation}
where $C_T$ is a  temperature-dependent
constant which turns out to be
\begin{equation}
\frac{C_T}{\kappa_0 k_{\rm BZ}} =
\sqrt{{\tilde{T}}/ {2 \pi}} \,.            \label{90}
\end{equation}
The symbol $ \tilde{T}$
denotes the dimensionless temperature
$ \tilde{T}\equiv  \tilde{\mu} k_B T/ (\kappa_0 k_{\rm BZ})^2 $.

The second coefficient
 $  \kappa_{\scriptstyle \Sigma} $ is determined
as follows. We assume that the
truncated
small-$k$ expansion
(\ref{@SIGEX}) can be used for all $k$
in the Brillouin zone, implying that
the inverse Green function has the form
\begin{eqnarray}
 G^{-1}({\bf k})
\approx C_T k^3 + \kappa_{r}k^4 ,
\label{@AAPP}\end{eqnarray}
with  $
 \kappa_r \equiv \kappa_0+\kappa_{\scriptstyle   \Sigma}$.
We shall see below that this
assumption is indeed
justified.
At low temperature where
  $  C_T \ll \kappa_r k_{\rm BZ} $ we determine
$ \kappa_r $ \mn{explain better}
by
inserting (\ref{@AAPP}) into
(\ref{80})
and evaluating the integral
for $ \Sigma (k)$
at the momentum $ k= C_T/ \kappa_r  $.
This momentum regime is most relevant
in the integrals over $ G $ which we have to calculate in the following
in order to determine the generalized Lindemann parameters.
Moreover, we will show below that
(\ref{80}) is then justified in good approximation
for momenta even in the whole Brioullin zone.
At higher temperatures where
$  C_T \gg \kappa_r
k_{\rm BZ} $, we determine
$ \kappa_r $ by integrating (\ref{80}) at $ k = k_{\rm BZ} $.
In both temperature regimes we
carry the momentum integrations up to
 $ k = k_{\rm BZ} $, and obtain
in either
case a
quadratic equation for $ \kappa_r $,
solved by
\begin{equation}
 \frac{\kappa_r}{\kappa_0}
\approx \left\{
\begin{array} {c c c}
  \frac{ 3 \tilde{T}}{8 \pi }
\left(
1- \sqrt{1- \frac{15}{16 \pi} \tilde{T}}\right)^{-1}   &  \mbox{for} &
C_T \ll \kappa_r k_{BZ},   \\
1- \frac{3}{4 \sqrt{2 \pi} }
\sqrt{\tilde{T}}   & \mbox{for} &
 C_T \gg \kappa_r k_{BZ}.
\end{array}
\right.
       \label{100}
\end{equation}
Our approximations are justified
in Fig.~1
showing in the main plot
the quantity $G^{-1}(k)\equiv  \kappa_r k^4+
C_T k^3  $ divided by the sum of $ \kappa_0 k^4 $ and the
numerically integrated right-hand side
of the self-energy function
(\ref{80}).
The numbers on the curves
 are  the various dimensionless temperatures $ \tilde{T} $.
 Observing that
the values of these curves are almost
constant and equal to unity
confirms
that $ G(k) $  of Eq.~(\ref{@AAPP})
indeed fulfills almost exactly the SCBA equation (\ref{80}).
 The inset of Fig.~1 shows $ \kappa_r / \kappa_0 $
as a function of the dimensionless temperature $ \tilde{T} $
calculated either by
(\ref{100}), corresponding in the figure
to the (green) solid and dashed curves,
 or by the determination of
$ \kappa_r / \kappa_0 $ by numerical integration of
the right-hand side
of (\ref{80}) ((blue) dashed-dotted  curve).
The kink in this curve corresponds to parameter values where
$ C_T/ \kappa_r k_{BZ} =1  $. Note that for graphene and SWNTs we have
$ \tilde{T}_m \approx 1.34$ at the melting point $ T_m  \approx 5000\,$K.

Next we calculate the expectation value
$ \langle \nabla_i f \nabla_j f \rangle $ where the average is taken
with respect to the Gibbs measure of the
Hamiltonian (\ref{30}) or (\ref{60}), respectively. In
SCBA, this leads to
\begin{align}
  \langle \nabla_i f \nabla_j f \rangle_T& \approx
\frac{1}{(2 \pi)^2} \int\limits_{\rm BZ} d^2 k \; k_i k_j G(k)
  \nonumber
   \\
& =
 \delta_{ij} \frac{\tilde{T}}{4 \pi}
\frac{\kappa_0 k^2_{\rm BZ}}{ \tilde{\mu}} \frac{\kappa_0} {\kappa_r}
\ln\left( 1+ \frac{\kappa_r k_{BZ}}{C_T}\right) \,.  \label{120}
\end{align}
Recalling (\ref{40})
we observe that
the
strain
in the $xy$-plane  is on the average equal to the negative of (\ref{120}):
 $ \langle \nabla_j u_i+\nabla_i u_j \rangle _T=
- \langle \nabla_i f \nabla_j
f \rangle_T $, implying that the self-induced stress
due to thermal out-of-plane
fluctuations vanishes.

  \begin{figure}
\begin{center}
\includegraphics[height=7cm,width=7.8cm]{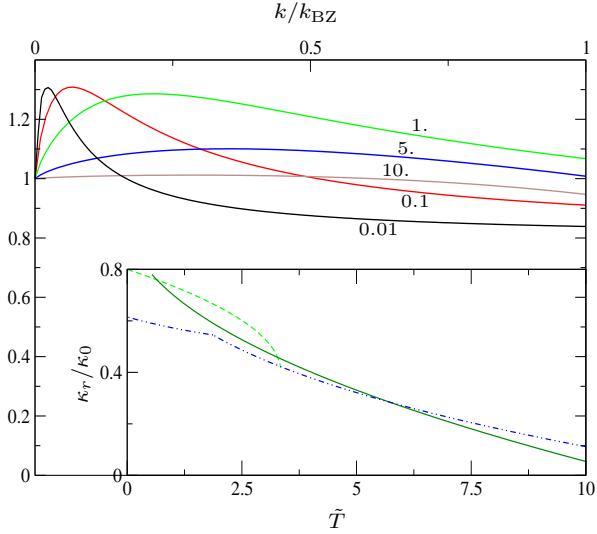}
\end{center}
\hspace*{-0.2cm}
\caption{(Color online)
Upper figure shows $G^{-1}(k)\approx C_T k^3 +\kappa_r k^4 $
divided by the sum of $ \kappa_0 k^4 $ and the
numerical integrated right-hand side of the self-energy function
(\ref{80}) where we used $ G^{-1}(k)= C_T k^3 + \kappa_r  k^4 $.
The numbers at the curves are
the dimensionless temperatures $ \tilde{T} $.
The inset shows
the
$ \tilde{T} $-behavior of  $ \kappa_r / \kappa_0 $.
The low-temperature dashed curve represents
Eq.~(\ref{100}) in the regime
$ C_T \ll \kappa_r k_{BZ} $,
the high-temperature solid curve
in the regime
$ C_T \gg \kappa_r k_{BZ} $.
The dashed-dotted curve
pictures
the ratio  $  \kappa_r / \kappa_0 $
obtained from
(\ref{80})
 by numerical
integration
of the right-hand side
with $ G^{-1}(k)= C_T k^3 + \kappa_r  k^4 $.
 }
\end{figure}

\subsection{Lindemann numbers}

We are no prepared to
calculate $( {\cal L}^c_1)^2 $ of  Eq.~(\ref{50}) by differentiating the
partition function of the elastic Hamiltonian $ H_{\rm el} $ twice
with respect to
the stress source  $ \sigma_{ij} $, and
setting
$ \sigma_{ij}=0 $ at the end.
Going over to the effective Hamiltonian
(\ref{60}) we obtain
two contributions $ ({\cal L}^c_1)^2= ({\cal L}^{s,2D}_{1})^2+
{\cal L}^2_{z1} $
where the first is the square of the Lindemann number
(\ref{10}) for the 2D hexagonal
solid
given by
\begin{align}
& ({\cal L}^{s,2D}_{1})^2=
\frac{1}{2}
 ( \langle  \nabla_i  u_m   \nabla_i  u_m \rangle_{2D} -
\langle  \nabla_i  u_m   \rangle_{2D}  \langle
\nabla_i     u_m \rangle_{2D})              \nonumber \\
& \qquad \approx \frac{k_B T}{\mu} \, \frac{3-\nu}{8} \, k^2_{\rm BZ} \,.
\label{130}
\end{align}
The average $ \langle \cdots \rangle_{2D} $ is calculated with respect
to the Gibbs
measure of the Hamiltonian (\ref{30}) with $ f =0 $ corresponding to the
2D crystal.
This contribution
to $({\cal L}^c_1)^2$
comes from the derivate of $ H_{\sigma \sigma } $ in Eq.~(\ref{60})
with respect to $ \sigma_{ij} $.
The second contribution
 $ {\cal L}^2_{z1} $ has its origin in the derivates
of the $ h_{h} $ term
in (\ref{60})  with respect
to $ \sigma_{ij} $, and is found to be
\begin{equation}
 {\cal L}^2_{z1}  =    (1 +\nu)^2 \frac{ 1 }{2 (2 \pi)^2}
\int\limits_{\rm BZ}  d^2 q \frac{1}{q^4}
\left(q_i q_j \! -\! \delta_{ij}
\frac{\nu}{1\!+\!\nu} q^2\right)^2\! I(q)  \label{140}
\end{equation}
where $ I(q)   $  is
the Fourier transform of the $ h_f^4 $ correlator
$
\langle h_f(x)  h_f(x') \rangle_T -
\langle h_f(x) \rangle_T \langle h_f(x') \rangle _T$.
Within self-consistent Born approximation we have
\begin{equation}
I(q)= \frac{(k_B T)^2}{2 (2 \pi)^2} \int\limits_{\rm BZ}  d^2k
 \frac{({\bf k} \times {\bf q})^4}{
q^4} G({\bf k}+{\bf q}) G({\bf k}) \, \Lambda({\bf k}+ {\bf q}, {\bf k}).
\label{150}
\end{equation}
The vertex correction factor $ \Lambda({\bf k}+ {\bf q}, {\bf k}) $ is
required within the SCBA by charge-current conservation.\mn{conservation of what?}
We first calculate (\ref{150}) in the lowest approximation
 $ \Lambda \approx 1 $,  to be justified
below. By using the analytic approximation
(\ref{@AAPP})
we obtain for (\ref{140}) with (\ref{150}) by integration
\cite{Smirnov1}
\begin{equation}
 \! \! {\cal L}^2_{z1}  \! \!
\approx \! \! \left\{ \! \!
\begin{array} {l c c}
  \frac{3(1+\tilde{\nu})^2 }{32 (2 \pi)^2 }
\tilde{T}^2 \left(\frac{ \kappa^2_0 k^2_{BZ}}{\kappa_r \tilde{\mu}} \right)^2
 \!\ln\left( \frac{\kappa_r k_{BZ}}{C_T} \right),\!\!
  &  \!\! \! \!\!\mbox{} &
~C_T \ll \kappa_r k_{BZ}   ,\\
 0.3 \frac{3 (1+\tilde{\nu})^2 }{32 (2 \pi)^2 }
\tilde{T}^2 \left(\frac{ \kappa_0 k^2_{BZ}}{\tilde{\mu}} \right)^2
\!\left(\frac{ \kappa_0 k_{BZ}}{C_T} \right)^2\!\!,\!\!
 & \! \! \!\!\mbox{} &
\! ~C_T \gg \kappa_r k_{BZ}.
\end{array}
\right.
       \label{160}
\end{equation}
 with $ (1+ \tilde{\nu})^2 =
(1+\nu)^2 \{ 1- 2 [\nu/(1+\nu)- \nu^2/(1+\nu)^2]\}$.

Our results depend strongly on the number
of the 2D Young modulus $E\equiv 2 \tilde{\mu} $.
The literature gives a broad range of possible
 $\tilde{\mu} $-values
(see for example \cite{Hsieh1,Dereli1} and references therein)
which makes
the comparison
of our results  with experiment
non-straightforward.
It was shown by Hsieh {\it et al.} \cite{Hsieh1}
using a molecular
dynamics simulation
that the
 Young modulus of (5,5) SWNTs
is softened
near the melting temperature \mn{???? check and explain better}
to around $ 70 \%$ of
 the $T=0$ -value, in agreement with Dereli {\it et al.}
 \cite{Dereli1} where a simulation of the larger
(10,10) SWNT was carried out. \mn{is this what you want to say?}
This value for the temperature reduction is in accordance
with the
temperature reduction of the 2D Young modulus in Wigner crystals
\cite{Morf1} at melting determined by computer simulation.
This can be generalized by theoretical
arguments to softening expressions of elastic moduli \cite{Dietel1,GFCM2}
for 2D crystals in general.
The $ T=0 $ -value for the (5,5) SWNT determined by
Hsieh {\it et al.} \cite{Hsieh1}
is $ E \approx 660 N/m $
and lies at the upper end
of existing Young moduli in the literature.
On the other hand the simulation of Dereli {\it et al.}
for the (10,10) SWNT results in a much lower value at room temperature
of around $ E \approx 140 N/m $ where it should not
much differ to its $ T=0 $ -value \cite{Hsieh1} lying at the
lower end of existing Young moduli for SWNTs in the literature.
One should compare this value
with the value $ E  \approx 440N/m $ found by Hsieh \cite{Hsieh1}
for the (17,0) tube taking into account that the
Young modulus depends only on the
diameter of the tube and not the helicity \cite{Hsieh1} in first approximation.
The origin of these discrepancies in the Young modulus values
shown in the literature in general is not clear yet.

To compare our analytic results with
the simulation results of Zhang {\it et al.} \cite{Zhang1}
we shall use in the following the $ T=0 $ -value
$ E \approx 350 N/m $ for the (5,5) SWNT
which is in the immediate proximity of
several simulations (see \cite{Huang2} and references therein)
and experimental values \cite{Lu1}.
The associated softened 2D Young modulus for (5,5) SWNTs
is thus $ E \approx 245 N/m $,
 which will
be used in the rest of the paper.
The remaining parameters are less significant
for our results. We shall
take
 $ \nu \approx 0.14 $,
$ \kappa_0 \approx 6.24^{-1} 10^{-18} {\rm Nm} $, $ k^2_{\rm BZ} \approx
2.46 \cdot 10^{20}/ {\rm m}^2 $
which are
typical for  SWNTs and graphene.
Inserting these
material parameters
we obtain for the melting temperature
\cite{Zhang1} $T_m = 5000\,$K a
value $ {\cal L}^2_z \approx  5.6 \cdot 10^{-4} $.
The contribution
(\ref{130}), on the other hand, adds to this
$ ({\cal L}^{s,2D}_{1})^2 \approx  8.9 \cdot 10^{-3}
$ at $ T_m \approx 5000$\,K
so that the modified Lindemann number
 $ {\cal L}^c_1 $
is $ \approx {\cal L}^{s,2D}_{1}  \approx 0.09 $.
This lies
in the
the range of values
$  {\cal L}^c_2 \approx 0.03 -0.07 $ obtained by numerical simulation
\cite{Zhang1}.

Our calculation shows that
the abrupt increase of $ {\cal L}^c_1 $ is
a meaningful criterium for the determination of the melting point.
At the melting point,
the
in-plane
shear modulus $ \mu $ will drop
to zero,  where
according to
Eqs.~(\ref{130}) and (\ref{160}),
$ {\cal L}^{s,2D}_{1}$ goes to infinity.

Next we calculate the 3D form of (\ref{10})
$ ({\cal L}^{s,3D}_1)^2 = ({\cal L}^{s,2D}_1)^2 +
{\cal L}_{z1}^2 +  {\cal L}_{z2}^2+ \frac{1}{2} \langle \nabla_i f \nabla_i f
\rangle_T  $ where the last term is due
to nearest-neighbor out-of-plane fluctuations given by (\ref{120}).
The first three terms measure
in-plane fluctuations where
$ {\cal L}^2_{z2} $ is given by the momentum integral
\begin{align}
&  {\cal L}^2_{z2}  =   - 2 (1 +\nu) \frac{ 1 }{2 (2 \pi)^2}
\int\limits_{\rm BZ}  d^2 q \frac{1}{q^2}
\bigg[(2\delta_{ij}-1)q_i q_j \!      \nonumber \\
&  -\! \delta_{ij}
\frac{\nu}{1\!+\!\nu} q^2\bigg]\,  I^1_{ij}(q)
+ \frac{ 1 }{2 (2 \pi)^2} \int\limits_{\rm BZ}  d^2 q
\,  I^2_{ij,ij}(q)  \,.
 \label{165}
\end{align}
The functions $ I^1_{ij} (q)$
and $ I^2_{ij,ij} (q)$
denote
the Fourier transforms of the expectation values
$$\!\!\!\frac{1}{2}\left[
\langle \nabla_i f(x)   \nabla_j f(x)  h_f(x')  \rangle_T
  -
\langle \nabla_i f(x)   \nabla_j f(x) \rangle_T \langle h_f(x') \rangle _T\right]  $$
and
\begin{eqnarray}
\!\!\!\!\!\!\!\!\!\!\!\frac{1}{4}\big[ \langle \nabla_i f(x)   \nabla_j f(x)  \nabla'_i f(x')   \nabla'_j f(x')
 \rangle_T ~~~~~~~~~~~~\nonumber \\ -
\langle \nabla_i f(x)   \nabla_j f(x) \rangle_T \langle
\nabla'_i f(x')   \nabla'_j f(x') \rangle _T\big] ,
\nonumber \label{@}\end{eqnarray}
respectively.
The contribution
$ {\cal L}_{z2}^2 $ is calculated in the same way as
$ {\cal L}_{z1}^2 $ \cite{Smirnov1},
yielding\mn{where is ${\cal L}_{z2}^2$ defined?}
\begin{equation}
 \! \! {\cal L}^2_{z2}  \! \!
\approx \! \! \left\{ \! \!
\begin{array} {l c c}
  \frac{3}{16 (2 \pi)^2 }
\tilde{T}^2 \left(\frac{ \kappa^2_0 k^2_{BZ}}{\kappa_r \tilde{\mu}} \right)^2
 \ln^2\left( \frac{\kappa_r k_{BZ}}{C_T} \right),\,
  &  \! \! \mbox{} &
C_T \ll \kappa_r k_{BZ}   ,\\
 \frac{3 (1+\tilde{\tilde{\nu}})}{16 (2 \pi)^2 }
\tilde{T}^2 \left(\frac{ \kappa_0 k^2_{BZ}}{\tilde{\mu}} \right)^2
\left(\frac{ \kappa_0 k_{BZ}}{C_T} \right)^2\!,\,
 & \! \! \mbox{} &
 C_T \gg \kappa_r k_{BZ}.
\end{array}
\right.
       \label{170}
\end{equation}
Here we have used the abbreviation
$ (1 + \tilde{\tilde{\nu}})\equiv  1+(1+\nu)[-0.31+0.25 \, \nu/(1+\nu)] $.\mn{wo kommen die krummen Zahlen her. Remark machen!}
Using the material  parameters
of SWNTs and graphene given above,
we find
 that the main contribution
 to $ {\cal L}^{s,3D}_1 $ comes from
the out-of-plane fluctuations and is given by
$ {\cal L}^{s,3D}_1 \approx  (\langle \nabla_i f \nabla_i f
\rangle_T/2)^{1/2}   \approx 0.22 $
at the melting point $ T_m \approx 5000\,$K. Thus, we find that
the out-of-plane fluctuations $ (\langle \nabla_i f \nabla_i f
\rangle_T/2)^{1/2} $ are even larger than the dominant contribution
to the in-plane fluctuations $ {\cal L}_1^{s,2D} $. By comparing the
temperature dependence (\ref{120}) with (\ref{130}) we obtain that
this is even the case for smaller temperatures.
Furthermore, we
realize that
in contrast to the Lindemann number ${\cal L}^c_1 $,
the abrupt increase of
the Lindemann number $ {\cal L}^{s,3D}_1 $ gives no
good signal for
the
melting point of a solid membrane. The reason
is that the vanishing
elastic shear modulus $ \mu $ at melting
contributes in two ways to the dominant fluctuation term
$ \langle \nabla_i f \nabla_i f
\rangle_T/2  $ (\ref{120}) but neither of them changes
this value much at melting.
First, the out-of-plane fluctuations
depend on $\mu$  via $ \kappa_r / \kappa_0 $ and remains
finite for
$ \mu \to 0 $, and second they depend
pick up
logarithmic dependence on $\mu$ from
$ C_T/ \kappa_0 k_{\rm BZ} $.
\mn{IS that OK}

Consider  now
the higher-order
vertex corrections collected in  the factor
$\Lambda({\bf k}+ {\bf q}, {\bf k}) $
in Eq.~(\ref{150}).
First we note that
for $ \Lambda\equiv 1 $ we obtain
 $ 2 \tilde{\mu} I(q)/k_B T < 3/8 $
in the dominant
integration regime of (\ref{140}) near
 $ q \approx C_T/\kappa $ for
$ C_T \ll \kappa_r k_{\rm BZ} $ and
near $ q \approx k_{\rm BZ} $ for
$ C_T \gg \kappa_r k_{\rm BZ} $. The factor $ 3/8 $ comes
mainly from the reduction of the
phase space integral by the projections $ P^T $ in the
polarisator. We expect that the $n$th order in $ \Lambda $
contributes roughly  with
a factor $ [2 \tilde{\mu} I(q)/k_B T]^n $
to $  I(q) $ in the dominant
integration regime of (\ref{140})
due to the additional phase space projection terms $ P^T $. \mn{correct?}
We have
checked this explicitly
by taking into account first-order corrections in the vertex
$ \Lambda $ in (\ref{150}).\mn{unverstaendlich}
A similar suppression of higher-order vertex correction contributions
occurs in $ I^1_{ij}(q) $ and $ I^2_{ijij}(q) $.

\subsection{Melting temperature}

Let us finally discuss the impact of the large
out-of-plane
fluctuations upon the melting temperature.
In Ref.~\onlinecite{Dietel1},
we have calculated
the melting temperature of a 2D triangular lattice
approximately
from the intersection
of high- and low-temperature expansion of the
free energies
associated with the Hamiltonian (\ref{30}) with
zero vertical displacements
$ f({\bf x})$. The transition
was caused by integer-valued
defect gauge fields accounting for the plastic deformations
of the crystal in the $xy$-plane.
These are coupled minimally to the
$xy$-displacement fields $u_i({\bf x}) $.
In that theory, the
melting temperature $T_m$ was found to obey the equation
\begin{equation}
\tilde{\beta}\equiv \frac{1}{k_B T_m}
\frac{\tilde{\mu}}{(2\pi)^2} \;  v_F \; \approx 0.6,
\label{180}
\end{equation}
where
$ v_F \equiv \sqrt{3} a^2 /2$ denotes the
$2D$-volume (area) of the fundamental cell.
 In SWNTs and
graphene, this
result is modified
by a factor
$S \; e^{-2W}$, where
$S$ is a
structure factor
and $ e^{-2W} $ a Debye-Waller-like factor
caused by the
out-of-plane fluctuations  $f({\bf x})\neq0$.
The
honeycomb lattice
of SWNTs contains
 two atoms
per triangular fundamental cell leading to
a structure factor $ S =1/2 $.
To estimate
the size of $e^{-2W}$
we observe that
in the defect melting model,
the defect gauge field appears
at a similar place
as
the vertical distortion  $ \nabla_i f  \nabla _j f /2 $
in the
Hamiltonian (\ref{30}).
Thus one can
immediately write down the Hamiltonian $ H_h $ of Eq.~(\ref{60})
coming from
defects. This leads
to the low-temperature expansion of the partition function.
In the
high-temperature expansion, there exist
a dual
stress representation
of the partition function \cite{GFCM2,Dietel1}.
In both low- and high-temperature representations,
the coupling
terms
between the defect fields or the
stress fields to the out-of-plane
fluctuations $ f({\bf x}) $ are
smaller than the
pure defect and stress term by approximately a factor
$ (4 \pi^2 \tilde{\beta})^2 {\cal L}^2_{z1}$ and $ (2\pi)^2 {\cal L}^2_{z2}
$, respectively.
When neglecting these small coupling terms
we find that the partition function
receives
a sizable correction factor
only in the the low-temperature approximation due
to the Fock energy of the Hamiltonian $ H_h $. The Hartree energy is missing
as a consequence of the open-boundary conditions
on the membrane \cite{Nelson2}.
From these considerations we obtain
\begin{align}
& W = \frac{v_F}{4} \frac{1}{(2 \pi)^2} \int\limits_{\rm BZ} d^2 k \, \Sigma(k) G(k)
\label{190}  \\
&  = \frac{1}{2} \frac{C_T}{\kappa_r k_{\rm BZ}}
\left[1- \frac{C_T}{\kappa_r k_{\rm BZ}} \ln\left( 1+
\frac{\kappa_r k_{BZ}}{C_T}\right)\right] \nonumber  \\
& + \frac{1}{2} \left(1\!-\! \frac{\kappa_0}{\kappa_r} \right)
\! \! \left[\frac{1}{2} -\frac{C_T}{\kappa_r k_{\rm BZ}}+
\left( \frac{C_T}{\kappa_r k_{\rm BZ}}\right)^2 \! \! \! \ln\left(\!  1\!+\!
\frac{\kappa_r k_{BZ}}{C_T}\! \right)\right].     \nonumber
\end{align}
Using the parameters above for (5,5) SWNTs we obtain $ W \approx 0.06 $ at
$ T \approx 5000\,$K. The factor $e^{-2W}
$ gives
thus only a small correction
to the melting temperature determined by (\ref{180}).
The explicit evaluation
of that equation
yields a melting temperature $ T_m \approx 8000\,$K ($ W \approx 0.075$),
somewhat larger than
the melting temperature $ T_m \approx 5000\,$K of
Zhang {\it et al.}  \cite{Zhang1} obtained by numerical simulation.

\section{Conclusion}
In this paper we have
calculated the fluctuations of solid membranes like graphene
and
single-walled carbon nanotubes
with the help of the
self-consistent Born-approximation.
Our results show that the
out-of plane fluctuations are much larger than the
in-plane fluctuations
even at low temperatures. Thus they may be expected to
have dramatic consequences
for the Lindemann numbers as well as the melting temperature
of solid membranes
in comparison to 2D crystals. Surprisingly, for the melting temperature 
this expectation was not confirmed.
The fluctuations was discussed by evaluating
the $3D$-version
$ {\cal L}^{s,3D}_1 $
 of the
Lindemann number (\ref{10}),
 originally
introduced to estimate the melting temperature
of 2D solids,
and the Lindemann number
${\cal L}^{c}_1$ defined in Eq.~(\ref{20}),
originally introduced in cluster physics.
We
observed that
a Lindemann criterium
based on ${\cal L}^{c}_1$ is
 more reliable
than that based on the former.
The associated
Lindemann number
is dominated by in-plane
fluctuations, in  contrast to the
former
which is dominated by the large out-of-plane
fluctuations.
By calculating, in addition,
the melting temperature
from a simple defect model of melting
for
single-walled carbon nanotubes
 and graphene (\ref{180}) we
observed  in contrast to the expectation, that the
melting temperature
depends
only very little on
the large out-of-plane fluctuations.

\acknowledgements
The authors acknowledge the support provided by Deutsche Forschungsgemeinschaft
under grant KL 256/42-2

\end{document}